# Magnetic properties of cobalt ultrathin film structures controlled by buffer layer roughness

Verbeno C.H.[1,a], Zázvorka J.[1,a,*], Nowak L.[1], Veis M.[1,*]

[1] *Institute of Physics, Faculty of Mathematics and Physics, Charles University, 12116, Prague, Czech Republic*
[a] joint first authors
*Corresponding authors

## Abstract

Growth optimization of multilayers is a topic of interest due to their unique physical properties, which can be different from those of their individual components. Systems containing magnetic materials, such as platinum-cobalt, have been studied because of their potential for technological applications, e.g. spintronics, magnetic storage and magnetic sensors. Since the magnetic properties of thin layers are strongly related to the growth parameters, the fine tuning of these parameters is necessary to produce multilayers with specific properties required in various applications. Here, an efficient approach to tune the coercive field of Co ultrathin films in the multilayer by varying the underlayer thickness is demonstrated. Using magnetron sputtering, we prepared multilayer systems of $Au_x/Pt_{5nm}/Co_{0.7nm}/Au_{5nm}$ with various thicknesses of Au underlayer. The surface morphology of $Au_x/Pt_{5nm}$ stack on which Co layer was deposited was studied by atomic force microscopy. We show the possibility to control the interfacial roughness by changing the Au underlayer thickness due to its island-like growth mechanism (Volmer-Weber mode). As the nominal thickness of Au increases, the islands grow in larger lateral size, resulting in a higher overall roughness of the layer surface. Magnetization measurements indicate a direct influence of the underlayer roughness on the coercivity of the multilayers by promoting additional magnetic anisotropy. With thickness of the Au layer up to 20 nm, we can change the coercive field in the range from ~200 Oe to ~1100 Oe, while remaining a nearly constant saturation magnetization. The use of Cu replacing Au underlayer in the same multilayers was also investigated, demonstrating the possibility of coercivity adjustment using different materials. The results are important for applications where the magnetic properties of multilayer structures based on Co thin films could be adjusted via buffer layer roughness engineering.

**Introduction**

Magnetic multilayers with perpendicular magnetic anisotropy (PMA) are currently the subject of interest due to their potential for technological applications, e.g. magnetic tunnel junctions (MTJs), magnetic random access memories (MRAMs), etc. Their resulting magnetic and transport properties are driven by several physical phenomena mediated by the multilayer composition, where combination of rare earth elements, transition metals and oxides are usually used to utilize mainly spin-orbit related interactions across interfaces. Therefore the properties of the functional multilayer stack can be effectively influenced during the growth process and for specific application purposes, optimization of the preparation procedure needs to be performed [1]. Individual layer thicknesses and the use of different magnetic materials can result in changes of magnetic parameters [2-4], such as saturation magnetization, magnetic coercivity and related quantities, e.g. magnetoresistance (GMR, TMR) ratios [5-7] and ferromagnetic resonance position and linewidth [8-10]. Magnetron sputtering is frequently used to grow metallic multilayers since it can achieve thickness of layers with precision in the order of Angstrom with well-defined interfaces. Recently, it has been shown that when using magnetic layers with ultra-low thicknesses, the interface of the layer and the substrate, or the seed layer, extensively influences the properties of the resulting multilayer [11-13]. Surface roughness and material intermixing are one of possible mechanisms behind the change of magnetic parameters of the magnetic layer, e.g. saturation magnetization and coercivity [14-16]. In the MTJ applications, the coercivity of the multilayer is desired to be large enough to ensure thermal stability upon magnetic state switching [17,18]. Other applications (e.g. magnetic field sensors) may require a specific coercivity or magnetic anisotropy. The modification of the interlayer after material growth using ion irradiation, laser irradiation was investigated previously [19-22]. However, these procedures require large facilities and rely of disrupting the material near the interface, which can influence not only coercivity, but unintentionally the saturation magnetization as well.

In this work, we perform a systematic study of a model Si/SiO$_2$/Au$_{x\,nm}$/Pt$_{5nm}$/Co$_{0.7nm}$/Au$_{5nm}$ magnetic multilayer system, where we vary the thickness of the Au buffer layer grown directly on the silicon substrate. By changing the buffer layer thickness, we modify the conveyed surface roughness of the platinum layer adjacent to the cobalt layer. Atomic force microscopy (AFM) is used to study the surface morphology of platinum layer on nominally identical samples without the magnetic layer and capping on top, prepared under the same sputtering condition. Magnetometry is used to obtain the information about saturation magnetization and coercivity of investigated samples with the different buffer layer thicknesses.

The experimental results show a good correlation between the Pt surface roughness, thickness of the Au layer and coercivity of the whole stack with almost no influence on saturation magnetization. The coercivity is varying considerably between 200 and 1100 Oe within 20 nm Au thickness range. This might be a big benefit for potential applications of magnetic multilayers in novel electronic devices. We also show that the use of the buffer layer is not constrained only to ferromagnetic Co, but it can be applied also for ferrimagnetic CoTb material as well.

**Experimental Procedures**

Two sets of samples with Au buffer layer of various thicknesses between 0 and 20 nm were prepared on naturally oxidized (100) undoped silicone substrates. The first one, $Au_x/Pt_{5nm}$ films to perform AFM measurements and, second one, $Au_x/Pt_{5nm}/Co_{0.7nm}/Au_{5nm}$ films to study the magnetic properties. In both cases x = 0, 0.5, 1, 1.5, 3, 5, 7, 10, 15, 20 nm. The samples were prepared by magnetron sputtering in ultrahigh vacuum (with base pressure around $10^{-9}$ mBar). Argon was used as the sputtering gas with a work pressure about $10^{-5}$ mBar. Silicon substrates were cleaned ex-situ by ultrasounds in an acetone and ethyl alcohol bath prior deposition. The top Au layer has capping function to avoid the oxidation of Co. All samples were grown at room temperature and the substrate holder was kept rotating at 30 rpm. Besides varying thickness of the Au layers, other deposition parameters of the samples did not vary. The same multilayer structures were also produced using Cu as a buffer layer instead Au (i.e. $Cu_x/Pt_{5nm}$ and $Cu_x/Pt_{5nm}/Co_{0.7nm}/Au_{5nm}$ films) with the purpose of investigating the possibility of coercivity adjustment using different materials.

The superficial morphology of the samples was characterized using an atomic force microscope (AFM, Witec Alpha 300) operating in contact mode. The AFM measurements were performed ex-situ immediately after the deposition in order to reduce possible effects of contamination and oxidation. Magnetic properties of the samples were investigated at 300K using a Physical Properties Measurement System (PPMS) by Quantum Design Inc. Sets of M(H) loops were recorded by vibrating sample magnetometer (VSM) experiments with the magnetic field applied either in-plane ($H_\parallel$) or out-of-plane ($H_\perp$) to the substrate`s surface. The measured magnetization data were corrected for the diamagnetic background of the substrates.

## Results and discussion

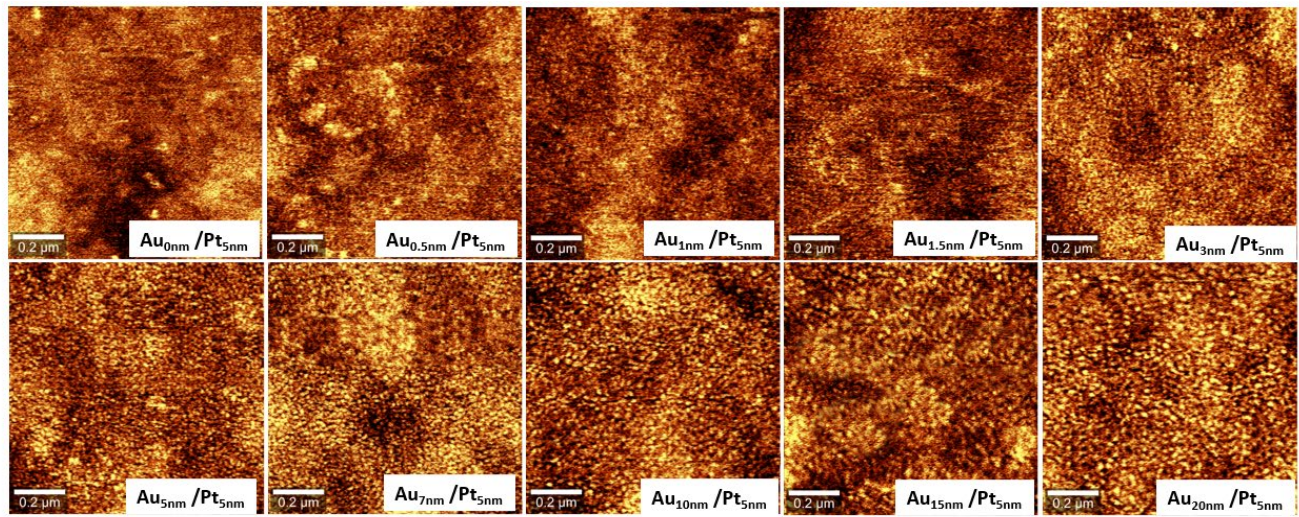

**Figure 1** – AFM images (4x4 um$^2$) of Au$_x$/Pt$_{5nm}$ films with various thickness of Au buffer layer.

**Fig. 1** shows AFM images obtained on Au$_x$/Pt$_{5nm}$ films with different thicknesses of Au. The samples present the formation of quasi-spherical nanoparticles (islands) on the surface, which is a characteristic of Volmer-Weber growth mode [23] (the surface energy of the Au/vacuum interface is lower than the Au/Pt interface). Also, the thinner samples (i.e., films with lower Au thickness) show smaller islands sizes in comparison with the thicker ones. This agrees with previous works on the polycrystalline thin films [24].

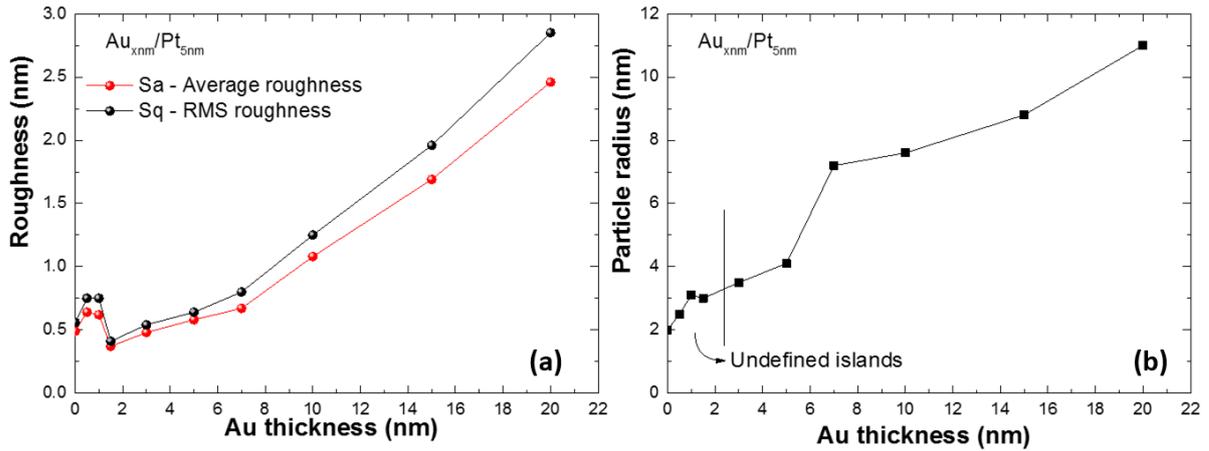

**Figure 2** – Roughness **(a)** and particle radius **(b)** as a function of Au thickness calculated from AFM measurements of $Au_{xnm}/Pt_{5nm}$ films.

The roughness and particle radius versus Au thickness calculated from AFM data are exhibited in **Fig. 2**. The average roughness ($S_a$) and rms roughness ($S_q$) increase sharply with the Au thickness, except for the samples with the thinnest Au layer (0.5 and 1 nm), where small roughness fluctuations are observed. The $S_a$ and $S_q$ parameters increased from ~0.5 nm to ~3 nm as the Au thickness increased from 1.5 to 20 nm. Ref. [14] describes the growth of thin films as a result of the competition between roughening by self-shadowing and the smoothening due to adatom surface diffusion. In this way, our results suggest that during this initial stage of growth (up to Au 20nm), self-shadowing is dominating. As observed in **Fig. 2b**, the particle sizes also increase with the increase of Au thickness indicating that the morphology of these films including the size and density of islands can be modified in the sputtering process. The islands grow in larger lateral size when the nominal Au thickness increases resulting in a higher overall roughness of the layer surface. This result shows the possibility to control Pt/Co interfacial roughness by changing the Au buffer layer thickness due to its island-like growth mechanism. It is worth to mention that the AFM measurements were performed ex-situ immediately after deposition to reduce potential oxidation and contamination effects. It was also confirmed that the main morphological characteristics are similar in various sample locations, indicating the produced samples are homogeneous.

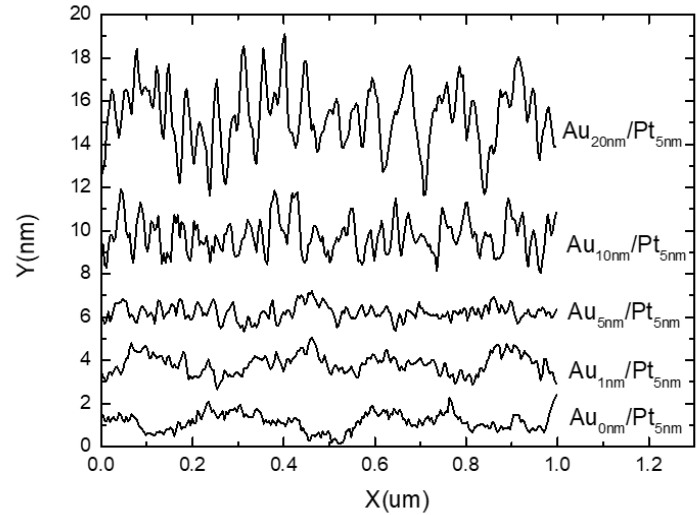

**Figure 3** – AFM profiles for $Au_x/Pt_{5nm}$ films with different Au thickness.

The evolution of AFM profiles for $Au_x/Pt_{5nm}$ films is shown in **Fig. 3**. In magnetron sputtering systems with oblique configuration, the self-shadowing effect is an important factor that can influence the superficial morphology of the films [25-28]. The increase in roughness leads to an intensification of the barriers (**Fig. 3**) where the incident Co adatoms are deposited. This results in a preferential deposition of Co on specific regions of the surface, which causes the formation of a discontinuous Co film. It is important to note that the roughness (varying from ~0.5 to ~3nm) becomes much larger than the Co layer thickness (0.7nm) as Au thickness increases (up to ~20nm). Therefore, the ultrathin Co film deposited on these surfaces will tend to become discontinuous.

To investigate the correlation between roughness and magnetic properties, $Au_x/Pt_{5nm}/Co_{0.7nm}/Au_{5nm}$ films were investigated by VSM with magnetic field applied either in plane ($H_{||}$) or out-of-plane ($H_\perp$) to surface. All the films exhibit an sharp magnetic state switching towards magnetic saturation in the $H_\perp$ configuration when compared to $H_{||}$ This indicates that the magnetic easy axis is predominantly oriented perpendicular to the sample surface, as demonstrated in **Fig. 4a** for a selected sample, $Au_{0.5nm}/Pt_{5nm}/Co_{0.7nm}/Au_{5nm}$. M(H) loops of all investigated samples recorded in out-of-plane configuration with various thickness of Au buffer layer are displayed in **Fig. 4b**. A clear increase of coercivity is visible with increasing Au buffer layer thickness.

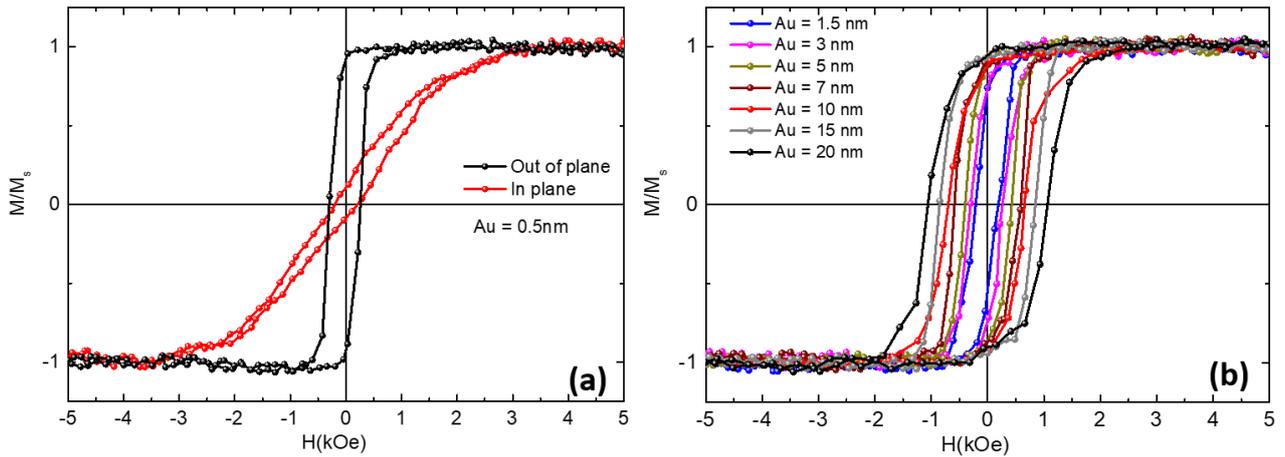

**Figure 4 – (a)** Comparison of M(H) loops recorded in both orientations out-of-plane and in plane for $Au_{0.5nm}/Pt_{5nm}/Co_{0.7nm}/Au_{5nm}$ sample. **(b)** Magnetic hysteresis loops for $Au_x/Pt_{5nm}/Co_{0.7nm}/Au_{5nm}$ samples with different thickness of Au measured in out-of-plane configuration

The coercivity, saturation magnetization and squareness ($M_r/M_s$) of the samples as functions of Au buffer layer thickness are plotted in **Fig. 5**. For in plane configuration, the coercivity ($H_c$) shows an increasing trend: the Au thickness changes $H_c$ from ~170 Oe to ~650 Oe. (**Fig. 5a**). For the out-of-plane $H_c$, two different regimes are observed. Firstly the $H_c$ decreases to a minimum value of ~200 Oe as Au thickness increases up to 1.5 nm, after that, $H_c$ continuously increases in the range from ~200 Oe to ~1100 Oe with Au thickness up to 20 nm. **Fig. 5a** also shows a comparison with the samples, where the buffer layer is in immediate proximity of the magnetic material, omitting the heavy metal layer. The comparison shows a similar trend with an offset in the amplitude to lower values, which will be addresses in detail in the discussion. From **Fig. 5b**. one can see that the squareness initially reduces to a minimum value of ~0.7 as the Au layer thickness increases up to 1.5 nm, however, further increasing to ~0.9 with the Au layer thicknesses (up to 20 nm) is observed. On the other hand, the saturation magnetization only slightly fluctuates around the value of 1300 emu/cm³.

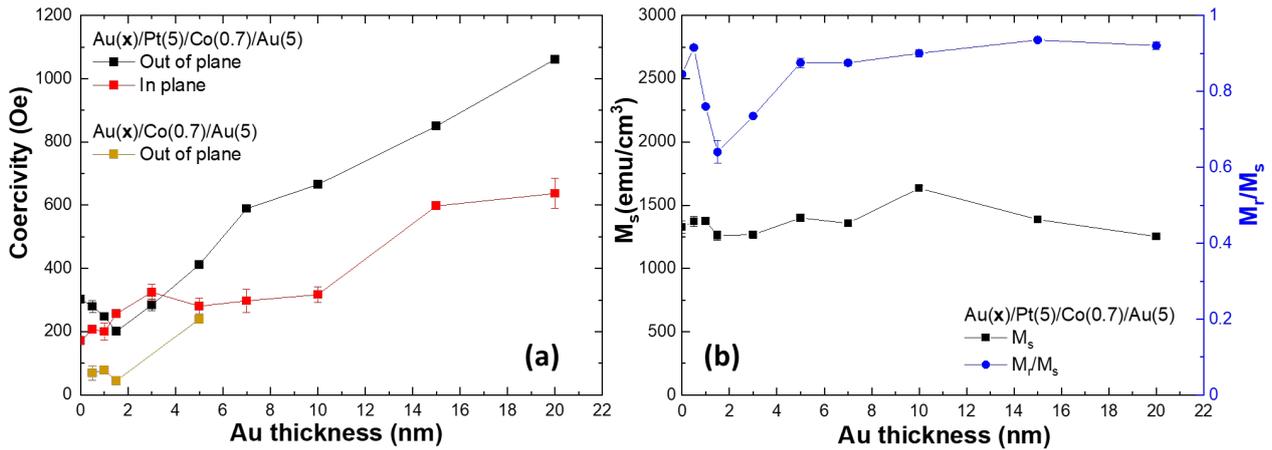

**Figure 5** – **(a)** Coercivity, **(b)** magnetization of saturation and squareness ($M_r/M_s$) as a function of Au thickness for $Au_x/Pt_{5nm}/Co_{0.7nm}/Au_{5nm}$ samples.

As the island-like Au buffer layer is becoming thicker, both roughness (**Fig. 2a**) and coercivity (**Fig. 5a**) increase dramatically with similar dependence (above $Au_{1.5nm}$). Also, there is a direct proportional relation between coercivity and roughness, as shown in **Fig. 6.** It is important to mention that when Au buffer layer is inserted in the multilayer (up to 1.5 nm) non-monotonous behavior of roughness is observed (**Fig. 2a**) while coercivity decreases (**Fig. 5a**). This distinct behavior can be explained considering that, at this regime of ultrathin layers, additional effects can be more influential to determine the magnetic properties, such as intermixing at interfaces [11,29,30].

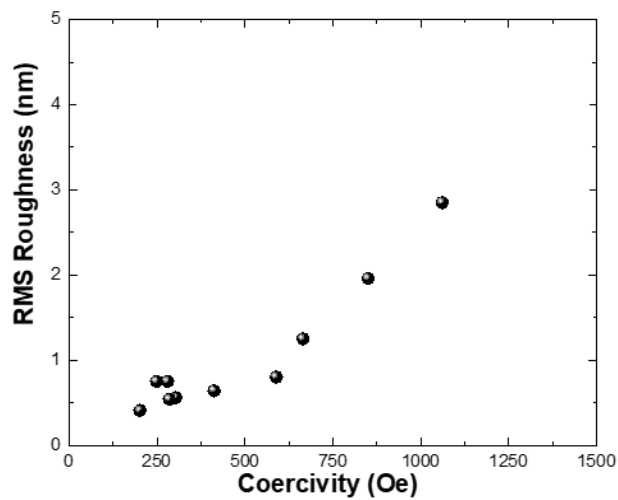

**Figure 6** – Correlation between coercivity and roughness of $Au_x/Pt_{5nm}/Co_{0.7nm}/Au_{5nm}$ samples.

Grain size and inner stress usually are the two main factors that can influence the coercivity of magnetic films [31-33]. Both can change the magnetic anisotropy constant and coercivity in a large scale. Prior works focusing on the effect of the grain size on coercivity reported that a larger grain size (associated with thicker magnetic layers) generally increases the coercivity [31,32]. However, in this work, the thickness of the magnetic layer is kept constant ($Co_{0.7nm}$) while the thickness of the buffer layer is altered. In this way, it is not expected to have considerable variations in the grain size of Co layer as a direct effect of growth of the magnetic material. Many other variables involved in sputtering process are known to greatly modify the final thin film properties, for example, substrate temperature and deposition power [30,34-36]. These possible factors were ruled out considering that all samples were produced under the same deposition conditions. Thus, the experiments were performed in a way to rule out any other phenomena that could explain the observed effects, for instance, tensile or compressive stress [35,36]. With the effect of grain size and inner stress excluded, the key influence on coercivity should come from the interfacial roughness, which can be modified by the deposition process.

Let us discuss in detail the roughness effect on the magnetic coercivity of the films. As showed in **Fig. 2 and 5**, the coercivity of $Au_x/Pt_{5nm}/Co_{0.7nm}/Au_{5nm}$ films change with Au buffer layer thickness in a similar way as surface roughness. Coercivity is a quantitative measure of the magnetic field required to reverse magnetization in the film and can be expressed as:

$$H_C = \frac{K_{eff}}{\mu_0 M_s}, \tag{1}$$

where $K_{eff}$ is the effective or total magnetic anisotropy energy, and $\mu_0 M_s$ is the saturation magnetization. Considering all anisotropy terms involved, the overall magnetic behavior of a given sample is described by an effective magnetic anisotropy, which consists of two main components [23]:

$$K_{eff} = K_V + \frac{K_S}{t}, \tag{2}$$

where $K_V$ is the volume anisotropy, which includes the shape and crystalline terms, and $K_S$ is the surface anisotropy that has an inverse proportionality dependence to the film thickness. According to equation (2), the influence of the surfaces (or interfaces, when in contact with another material) plays a major role in thinner films as $K_s$ becomes more dominant in the effective anisotropy of magnetic structures.

Then, considering that all samples were produced with an ultrathin layer of Co (0.7 nm thick), it is reasonable to expect a major contribution of $K_s$ on the magnetic properties. However, as discussed before, the Co layer (0.7 nm thick) thickness is kept constant throughout the study and following from the nearly constant saturation magnetization, the amount of deposited material is considered to remain constant as well. Deposition on very rough surfaces ($R_q$ values up to 3 nm) induces the formation of discontinuous interface, resulting in magnetic ripples and in domain wall pinning. This tend to hinder domain wall displacement and increases the coercivity [29,37]. While in the intermediate regime, where the magnetic material can be considered as continuous, increasing the surface roughness effectively increases the surface or area where the magnetic material comes in contact with the adjacent layer of heavy metal (in our case platinum), which can affect the magnetic layer itself through proximity effects, changing the magnetic parameters.

Zhao and coworkers [11] proposed a model to study the effect of surface roughness on coercivity of magnetic thin films. The authors showed that the roughness of an isotropic self-affine surface modifies the demagnetizing factors and induces changes in the magnetostatic energy differently for Bloch walls and Néel walls. The coercivity of a magnetic thin film caused by domain wall movement was explicitly written as (equation 31 in Ref. [11])

$$H_C = \frac{1}{2M_s}\left(\frac{A_{ex}\pi^2}{D_w\, t} + \frac{K_V D_w}{2t} + \frac{D_w t + 2D_w^2}{(t+D_w)^2}\pi M_s\right) \cdot \rho_{rms} \quad (3)$$

where $A_{ex}$ is the exchange constant, $D_w$ is the domain wall thickness, $M_s$ is the saturation magnetization and $\rho_{rms}$ is a dimensionless roughness. According to equation (3), for the same film thickness $t$ and constant saturation magnetization, the rougher the surface, the larger the coercivity, which agrees well with the experimental results presented in this work. This is inherently valid for the magnetic material itself with changing surface roughness. Comparison of the samples containing platinum as the heavy metal layer with samples without this layer (Fig. 5a) reveals similar Au thickness behaviour with the coercivity offset towards higher values for Pt, which can be a sign of proximity effects. These can be pronounced due to the higher interface area with higher roughness. Both presented mechanism will affect the change of the coercivity while preserving the saturation magnetisation of the magnetic layer.

The use of Cu instead of Au buffer layer in the same multilayer structure was also investigated. For this purpose, $Cu_x/Pt_{5nm}$ and $Cu_x/Pt_{5nm}/Co_{0.7nm}/Au_{5nm}$ samples were produced under the same growth conditions and studied using AFM and VSM measurements. The evolution of the surface morphology of $Cu_x/Pt_{5nm}$ samples presents similar behavior when compared to the Au buffer layer system: formation of islands on the surface that grow in larger lateral size with the increasing of Cu buffer layer thickness, resulting in a higher overall roughness, as observed in the AFM images of $Cu_{1.5nm}/Pt_{5nm}$ and $Cu_{20nm}/Pt_{5nm}$ samples (**Figs. 7a-b**). Also, the same correlation between coercivity and roughness is noticed since both parameters increase with the Cu buffer layer thickness (**Figs. 7c-d**). These results support our conclusion about the induced effective magnetic anisotropy due to higher interface area with higher surface roughness and, also, demonstrate the possibility of coercivity adjustment using different materials for applications where the magnetic properties of structures based on Co thin films could be controlled via buffer layer roughness engineering allowing the optimization of these devices.

From the application point of view, using platinum is desired as it has a strong spin coupling and can be used in detecting spin current. Therefore, the thickness of Au buffer layer turned out to be a crucial parameter for optimization of Pt/Co multilayer devices, where adequate PMA with well-defined magnetization switching behavior can be controlled through tunning of the magnetic coercivity.

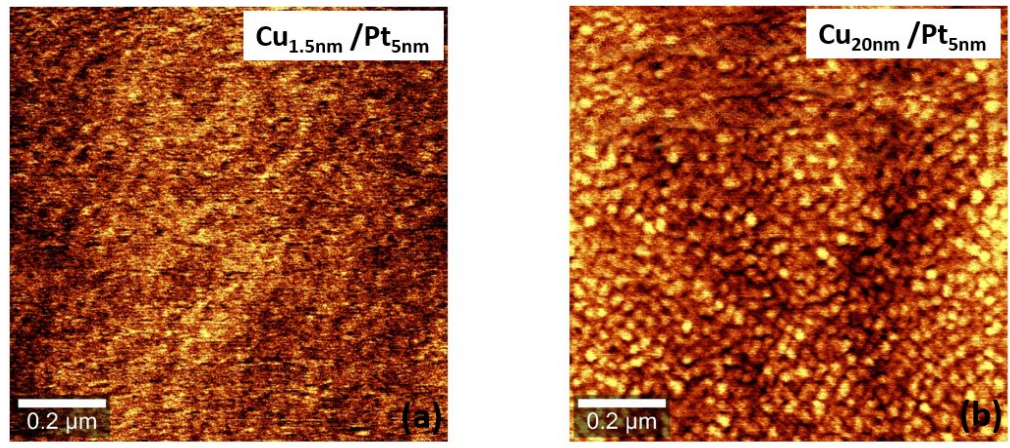

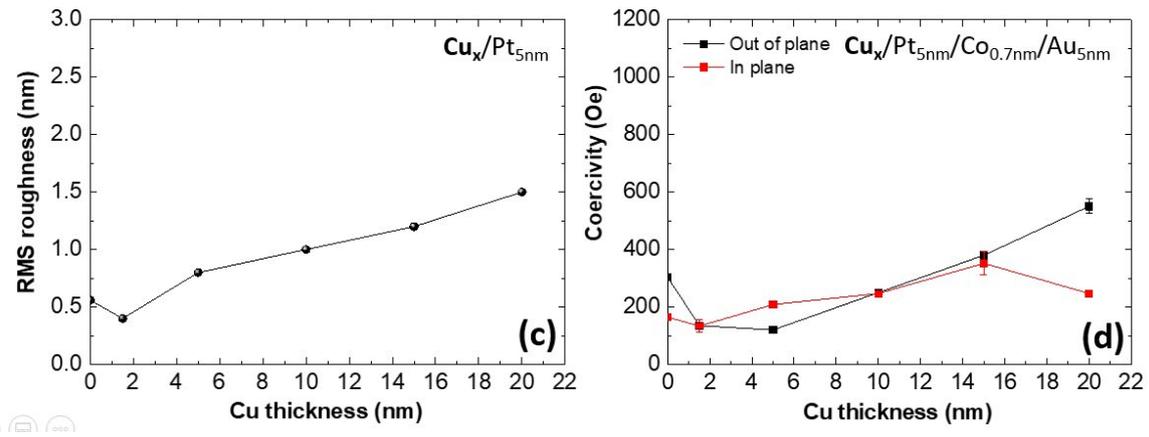

**Figure 7** – AFM images (4x4 um$^2$) of **(a)** Cu$_{1.5nm}$/Pt$_{5nm}$ and **(b)** Cu$_{20nm}$/Pt$_{5nm}$. **(c)** Cu thickness dependence of roughness for Cu$_x$/Pt$_{5nm}$ samples. **(d)** Coercivity values measured for Cu$_x$/Pt$_{5nm}$/Co$_{0.7nm}$/Au$_{5nm}$ samples.

To demonstrate that the presented approach is not limited only to ferromagnetic Co, we have exchanged Co with ferrimagnetic Co/Tb multilayer. **Fig. 8** presents hysteresis loops of two ferrimagnetic stacks consisted of Au$_x$/Pt$_{5nm}$/[Co$_{0.7nm}$/Tb$_{0.4nm}$]$_{x6}$/Au$_{5nm}$ with different thicknesses of Au buffer layer, 1.5nm and 20 nm. The coercivity increases with the Au thickness as expected, which shows the potential of layer roughness engineering method also for ferrimagnetic multilayers with larger number of repetitions.

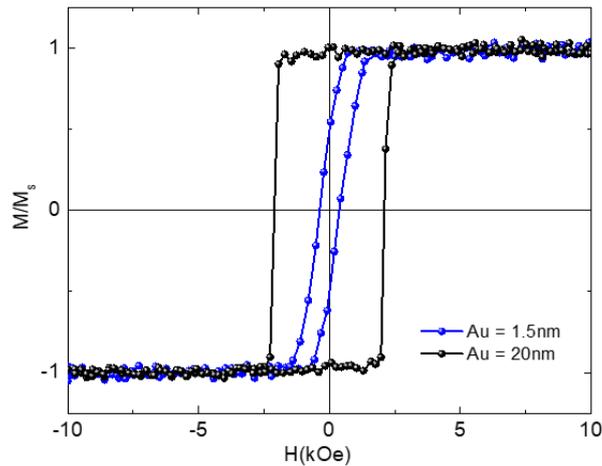

**Figure 8** – Comparison of out-of-plane hysteresis loops measured at 300K for two ferrimagnetic structures $Au_x/Pt_{5nm}/[Co_{0.7nm}/Tb_{0.4nm}]_{x6}/Au_{5nm}$ with different thickness of Au buffer layer.

## Conclusions

We investigated the interface roughness and magnetic properties of $Au_x/Pt_{5nm}/Co_{0.7nm}/Au_{5nm}$ multilayers as function of the Au buffer layer thickness using VSM and AFM techniques. The results showed that as the Au buffer layer becomes thicker, both roughness and coercivity increase significantly. Thus, the interface roughness, which can be controlled adjusting Au buffer layer thickness, proved to be a critical factor that influences the properties of magnetic films in a multilayer system containing platinum. The utilization of Cu instead of Au buffer layer in the same multilayer structure was also investigated (i.e. $Cu_x/Pt_{5nm}/Co_{0.7nm}/Au_{5nm}$ films) to demonstrate the possibility of coercivity adjustment using different materials as buffer layers. The experiments on Tb/Co samples validated the use of this approach also for ferrimagnetic multilayers. These results demonstrated an effective method to tune physical properties of magnetic multilayers with relatively high precision.

## Acknowledgements

This work was supported by the Charles University PRIMUS project (Grant No. PRIMUS/20/SCI/018) and the Ministry of Education, Youth and Sports of the Czech Republic from the OP RDE programme (Project No. MATFUN CZ.02.1.01/0.0/0.0/15_003/0000487). The authors thank the MGML laboratory in Prague for the help with magnetometry measurements.